\documentclass[12pt]{article}
\usepackage{euscript,amsmath,amssymb,latexsym}

\newcommand{\be}{\begin{equation}}
\newcommand{\ee}{\end{equation}}
\newcommand{\bea}{\begin{eqnarray}}
\newcommand{\eea}{\end{eqnarray}}

\begin{document}
\title{
\vspace{1cm} {\bf  Randomness in  Classical Mechanics and \\
Quantum Mechanics}
 }
\author{Igor V. Volovich
 \\
{\it  Steklov Mathematical Institute}
\\ {\it Gubkin St.8, 119991 Moscow, Russia}
\\ email:\,volovich@mi.ras.ru}

\date {~}
\maketitle

\begin{abstract}
The  Copenhagen interpretation of quantum mechanics assumes the
existence of the classical deterministic Newtonian world. We argue
that in fact  the Newton determinism in classical world does not
hold  and  in classical mechanics there is fundamental and
irreducible randomness. The classical Newtonian trajectory does not
have a direct physical meaning since arbitrary real numbers are not
observable. There are classical uncertainty relations: $\Delta q
>0$ and $\Delta p>0$, i.e. the uncertainty (errors of observation)
in the determination of coordinate and momentum is always positive
(non zero).

A ``functional" formulation of classical mechanics was suggested.
The fundamental equation of the microscopic dynamics in the
functional approach is not the Newton equation but the Liouville
equation for the distribution function of the single particle.
Solutions of the Liouville equation have the property of
delocalization which accounts for irreversibility. The Newton
equation in this approach appears as an approximate equation
describing the dynamics of the average values of the position and
momenta for not too long time intervals. Corrections to the Newton
trajectories are computed. An interpretation of quantum mechanics is
attempted in which both classical and quantum mechanics contain
fundamental randomness. Instead of an ensemble of events one
introduces an ensemble of observers.
\end{abstract}
\maketitle

\newpage

\section{Introduction}

In classical mechanics the motion of a point body is described by
the trajectory in the phase space, i.e. the values of the
coordinates and momenta as functions of time, which are solutions of
the equations of Newton or Hamilton, \cite{New}.

However,  this mathematical model is an idealization of the physical
process, rather far separated from reality. The physical body always
has the spatial dimensions, hence a mathematical point gives only an
approximate description of the physical body. The mathematical
notion of a trajectory does not have direct physical meaning, since
it uses  arbitrary real numbers, i.e. infinite decimal expansions,
while the observation is only possible, in the best case, of
rational numbers, and even them only with some error. Therefore, we
suggest a ``functional" approach to classical mechanics, which is
not starting from Newton's equation, but with the Liouville
equation. This approach can help to explain the infamous time
irreversibility problem, see, for example \cite{Bol1} - \cite{Koz1}.

The conventional widely used concept of the microscopic state of the
system at some  moment in time  as the point in phase space, as well
as the notion of trajectory and the microscopic equations of motion
have no direct physical meaning, since arbitrary real numbers not
observable (observable physical quantities are only presented by
rational numbers, cf. the discussion of concepts of space and time
in \cite{ Vol1} - \cite{DKKV}).

In the  functional approach \cite {VolIrr} the physical meaning is
attached not to a single trajectory but only to a ``beam" of
trajectories, or the distribution function on phase space.
Individual trajectories are not observable, they could be considered
as ``hidden variables", if one uses the quantum mechanical notions.

The fundamental equation of the microscopic dynamics of the
functional probabilistic approach is not Newton's equation, but a
Liouville equation for distribution function. It is well known that
the Liouville equation is used in statistical mechanics for
description of the motions of gas. Let us stress that we shall use
the Liouville equation even for  description of a single particle in
the empty space.

There are many discussions of the time irreversibility problem, i.e.
the problem of how to explain the irreversible behaviour of
macroscopic systems from the time-symmetric microscopic laws.  The
problem has been discussed by
 Boltzmann, Poincar\'e, Bogolyubov, Feynman and many other authors,
  \cite{Bol1} - \cite{Koz1}. Landau
and Lifshiz wrote about the principle of increasing entropy
  \cite{LL}: ``Currently it is not clear whether the
  law of increasing entropy  can be in principle derived from classical
  mechanics." Landau speculated that to explain
  the second law of thermodynamics one has to use quantum mechanical
  measurement arguments.

Although the Liouville equation is symmetric in relation to the
reversion of time, but his solutions have the property of {\it
delocalization}, that, generally speaking, can be interpreted as a
manifestation of irreversibility. It is understood that if at some
moment in time the distribution function describes a particle,
localized to a certain extent,  then over time the degree of
localization decreases, there is the spreading of distribution
function. Delocalization takes place even for a free particle in
infinite space, where there is no ergodicity and mixing. In a sense,
the functional formulation of microscopic dynamics  is irreversible
in time. Thus  the contradiction between microscopic reversibility
and macroscopic irreversibility of the dynamics disappears, since
both microscopic and macroscopic dynamics in the proposed approach
are irreversible.

In the functional approach to classical mechanics we do not derive
the statistical or chaotic properties of deterministic dynamics, but
we suggest that the Laplace's determinism  at the fundamental level
is absent not only in quantum, but also in classical mechanics.

We show that Newton's equation in the proposed approach appears as
an approximate equation describing the dynamics of the average
values of coordinates and momenta for not too long  time. We
calculate corrections to Newton's equation.

An interpretation of quantum mechanics is attempted in which both
classical and quantum mechanics contain fundamental randomness.
Instead of an ensemble of events one introduces an ensemble of
observers.

In the next section the fundamentals of the functional formulation
of classical and quantum mechanics are presented. Section 3 deals
with the free movement of particles and Newton's equation for the
average coordinates. Comparison with quantum mechanics is discussed
in Section 4. General comments on the Liouville and Newton equations
are given in section 5. Corrections to the Newton equation for a
nonlinear system are calculated in Section 6.  The dynamics of the
classical and quantum particle in a box and their interrelationships
are summarized in section 7.

\section{States and Observables in   Functional  Mechanics}

\subsection{Classical mechanics}

An {\it exact} derivation of the coordinate and momentum can not be
done, not only in quantum mechanics, where there is the Heisenberg
uncertainty relation, but also in classical mechanics. Always there
are some errors in setting the coordinates and momenta. There are
classical uncertainty relations: $\Delta q
>0$ and $\Delta p>0$, i.e. the uncertainty (errors of observation)
in the determination of coordinate and momentum is always positive
(non zero). The concept of arbitrary real numbers, given by the
infinite decimal series, is a mathematical idealization, such
numbers can not be measured in the experiment.

Consider the motion of a classical particle along a straight line in
the potential field. The general case of many particles in the
3-dimensional space is discussed below. Let $(q, p)$ be coordinates
on the plane $\mathbb{R}^2$  (phase space), $t\in\mathbb{R}$ is
time. The state of a classical particle at time $t$ will be
described by the function $\rho=\rho (q, p, t)$, it is the density
of the probability that the particle at time $t$ has the coordinate
$q$ and momentum $p$.

In \cite{TV2} it is given a construction of the probability density
function starting from the directly observable quantities, i.e., the
results of measurements, which are rational numbers.

\subsection{Classical and quantum mechanics}
Note that the description of a mechanical system with the help of
probability distribution function $\rho=\rho (q, p, t)$ does not
necessarily mean that we are dealing with a set of identically
prepared ensemble of particles. Usually in probability theory one
considers  an ensemble of events  and a sample space \cite{Khr2}.
But we can use the description with the function $\rho=\rho (q, p,
t)$ also for individual bodies, such as planets in astronomy (the
phase space in this case the 6-dimensional). In this case one can
think on the ``ensemble" of different astronomers which observe the
planet. It might be that there is only one ``intelligent" observer
and an ``ensemble" of different scenario of behaviour  of a given
object  in such a way that one can deal with  an individual quantum
phenomenon. From this point of view there is no difference between
``Einstein`s moon" and ``Heisenberg`s electron".  Actually, it is
implicitly always dealt with the function $\rho=\rho (q, p, t)$
which takes into account the inherent uncertainty in the coordinates
and momentum of the body.

The wave function in quantum mechanics $\psi =\psi (q,t)$ or the
density operator actually depends not only from time $t$ and
position $q$ but also from other parameters such as the form of the
potential field and the length of the box as well as from the mass,
charge and the Planck constant. Denote these parameters by $\xi$.
Some of these parameters can be called the ``contextual" variables.
We have the wave function $\psi =\psi (q,t;\xi)$. In the functional
formulation of quantum mechanics we introduce a distribution $\sigma
(\xi)$ which describes the uncertainty in the derivation of these
parameters. To get  observed quantities we have to evaluate the
average value of $\psi$ or $|\psi|^2$ with $\sigma (\xi)$. More
discussions of functional quantum mechanics will be presented in a
separate work. Note that similar distribution $\sigma (\xi)$ we have
to introduce already in classical functional mechanics.

The specific type of function $\rho$ depends on the method of
preparation of the state of a classical particle at the initial time
and the type of potential field. When $\rho=\rho (q, p, t)$ has
sharp peaks at $q = q_0$ and $p = p_0$, we say that the particle has
the approximate values of coordinate and momentum $q_0$ and $p_0$.

In the functional approach to classical mechanics the concept of
precise trajectory of a particle is absent, the fundamental concept
is a distribution function $\rho=\rho (q, p, t)$ and
$\delta$-function as a distribution function is not allowed.

We assume that the continuously differentiable and integrable
function $\rho=\rho (q, p, t)$ satisfies the conditions:
\begin{equation}\label{rho}
\rho \geq 0,~~\int_{\mathbb{R}^2}\rho (q,p,t)dqdp=1,~ t\in
\mathbb{R}\,.
\end{equation}
The motion of particles in the functional approach is not described
directly by the Newton (Hamilton) equation. Newton's equation in the
functional approach is an approximate equation for the average
coordinates of the particles, and for non-linear dynamics  there are
corrections to the Newton equations.

If $f = f (q, p)$  is a function on phase space, the average value
of $f$ at time $t$ is given by the integral
\begin{equation}\label{int1}
\overline{f}(t)=\int f(q,p)\rho (q,p,t)dqdp\,.
\end{equation}
In a sense we are dealing with a random process $\xi(t)$ with values
in the phase space.

\subsection{Basic equation for a single particle}

Motion of a point  body along a straight line in the potential field
will be described by the equation
\begin{equation}\label{Lio}
\frac{\partial \rho}{\partial t}=-\frac{p}{m}\frac{\partial
\rho}{\partial q}+\frac{\partial V(q)}{\partial q} \frac{\partial
\rho}{\partial p}\,.
\end{equation}
Here $V(q)$  is the potential field and mass $m>0$.

Equation (\ref{Lio}) looks like the Liouville equation which is used
in statistical physics to describe a gas of particles but here we
use it to describe a single particle.

If the distribution $\rho_0(q,p)$ for $t=0$ is known, we can
consider the Cauchy problem for the equation (\ref{Lio}):
\begin{equation}\label{cau}
\rho |_{t=0}=\rho_0(q,p)\,.
\end{equation}
Let us discuss the case when the initial distribution has the
Gaussian form:
\begin{equation}\label{gauss}
\rho_0 (q,p)=\frac{1}{\pi
ab}e^{-\frac{(q-q_0)^2}{a^2}}e^{-\frac{(p-p_0)^2}{b^2}}\,.
\end{equation}
At sufficiently small values of the parameters $a> 0$ and $b> 0$ the
particle has coordinate and momentum close to the $q_0$ and $p_0$.
For this distribution the average value of the coordinates and
momentum are:
\begin{equation}\label{mean1}
\overline {q}=\int q\rho_0 (q,p)dqdp=q_0\,,~~\overline{p}=\int
p\rho_0 (q,p)dqdp=p_0\,,
\end{equation}
and dispersion
\begin{equation}\label{disp1}
\Delta q^2=\overline{(q-\overline {q})^2}=\frac{1}{2}a^2,~~\Delta
p^2=\overline{(p-\overline {p})^2}=\frac{1}{2}b^2\,.
\end{equation}

\section{Free Motion} Consider first the case of the free motion of
the particle when $V=0$. In this case the equation (\ref{Lio}) has
the form
\begin{equation}\label{free}
\frac{\partial \rho}{\partial t}=-\frac{p}{m}\frac{\partial
\rho}{\partial q}
\end{equation}
and the solution of the Cauchy problem is
\begin{equation}\label{solfree} \rho (q,p,t)=\rho_0
(q-\frac{p}{m}t,p)\,.
\end{equation}
Using expressions (\ref{gauss}), (\ref{solfree}),
\begin{equation}\label{gauss2}
\rho (q,p,t)=\frac{1}{\pi
ab}\exp\{-\frac{(q-q_0-\frac{p}{m}t)^2}{a^2}-\frac{(p-p_0)^2}{b^2}\}\,,
\end{equation}
we get the time dependent distribution of coordinates:
\begin{equation}\label{coor1}
\rho_{c} (q,t)=\int \rho (q,p,t)dp=\frac{1}{\sqrt{\pi}
\sqrt{a^2+\frac{b^2t^2}{m^2}}}\exp\{-\frac{(q-q_0-
\frac{p_0}{m}t)^2}{(a^2+\frac{b^2t^2}{m^2})}\}\,,
\end{equation}
while the distribution of momenta is
\begin{equation}\label{mom1}
\rho_{m} (p,t)=\int \rho
(q,p,t)dq=\frac{1}{\sqrt{\pi}b}e^{-\frac{(p-p_0)^2}{b^2}}.
\end{equation}
Thus, for the free particle the distribution of the particle
momentum with the passage of time does not change, and the
distribution of the coordinates change. There is, as one says in
quantum mechanics, the spreading of the wave packet. From
(\ref{coor1}) it follows that the dispersion $\Delta q^2$ increases
with time:
\begin{equation}\label{disp2}
\Delta q^2(t)=\frac{1}{2}(a^2+\frac{b^2t^2}{m^2})\,.
\end{equation}
Even if the particle was arbitrarily well localized ($a^2$ is
arbitrarily small) at $t = 0$, then at sufficiently large times $t$
 the localization of the particle becomes meaningless, there is a
{\it delocalization} of the particle.

What role can play the Newton equation in the functional approach?
The average  coordinate for the free particle in the functional
approach satisfies the Newton equation. Indeed, the average
coordinate and momentum for the free particles have the form
\begin{equation}\label{coor2} \overline {q}(t)=\int
q\rho_{c} (q,t)dq=q_0+\frac{p_0}{m}t\,,~~\overline{p}(t)=\int
p\rho_m(p,t)dp=p_0\,.
\end{equation}
Note that in the functional mechanics the Newton equation for the
average  coordinates is obtained only for the free particle or for
quadratic Hamiltonians with a Gaussian initial distribution
function. For a more general case there are corrections to Newton's
equations, as discussed below.

\section{Comparison with Quantum Mechanics}
Compare the evolutions of Gaussian distribution functions in
functional classical mechanics and in quantum mechanics for the
motion of particles along a straight line. The scene of work for the
functional classical mechanics is $ L ^ 2 (\mathbb {R} ^ 2) $ (or $
L ^ 1 (\mathbb {R} ^ 2) $), and for quantum mechanics - $ L ^ 2
(\mathbb {R} ^ 1) $.

The Schrodinger equation for a free quantum particle on a line
reads:
\begin{equation}\label{quan}
i\hbar\frac{\partial \psi}{\partial
t}=-\frac{\hbar^2}{2m}\frac{\partial^2\psi}{\partial x^2}\,.
\end{equation}
Here $\psi=\psi(x,t)$  is the wave function and $\hbar$ is the
Planck constant. The density of the distribution function for the
Gaussian wave function has the form
\begin{equation}\label{quan2}
\rho_q(x,t)=|\psi (x,t)|^2=\frac{1}{\sqrt{\pi}
\sqrt{a^2+\frac{\hbar^2t^2}{a^2m^2}}}\exp\{-\frac{(x-x_0-
\frac{p_0}{m}t)^2}{(a^2+\frac{\hbar^2t^2}{a^2m^2})}\}\,.
\end{equation}
We find that the distribution functions in functional classical and
in quantum mechanics (\ref {coor1}) and (\ref {quan2}) coincide, if
we set
\begin{equation}\label{unc}
a^2 b^2=\hbar^2\,.
\end{equation}
If the condition (\ref {unc}) is satisfied then the Wigner function
$ W (q, p, t) $  for $ \psi $ corresponds to the classical
distribution function (\ref {gauss2})\,, $ W (q, p, t) = \rho (q, p,
t) $ \,.

Gaussian wave functions on the line are coherent or compressed
states. The compressed states on the interval are considered in
\cite {VT}.

\section {Liouville Equation and the Newton Equation}

 In the functional classical mechanics  the motion of a particle
 along the stright line is described by the
  Liouville equation (\ref {Lio}). A more general Liouville equation
  on the manifold $ \Gamma $ with coordinates
  $ x = (x^1 ,..., x^k) $ has the form
\begin{equation}\label{Lio2a}
\frac{\partial \rho}{\partial t}+\sum_{i=1}^k
\frac{\partial}{\partial x^i}(\rho v^i)=0\,.
\end{equation}
Here $\rho=\rho (x,t)$ is the density function and
$v=v(x)=(v^1,...,v^k)$ - vector field on $\Gamma$. The solution of
the Cauchy problem for the equation  (\ref{Lio2a}) with initial data
\begin{equation}\label{Lio3}
\rho|_{t=0}=\rho_0(x)
\end{equation}
might be written in the form
\begin{equation}\label{Lio4}
\rho (x,t)=\rho_0(\varphi_{-t}(x))\,.
\end{equation}
Here $\varphi_t(x)$ is a phase flow along the solutions of the
characteristic equation
\begin{equation}\label{char2}
\dot{x}=v(x)\,.
\end{equation}
A system in the phase space $\mathbb{R}^{2N}$ with coordinates
$q=(q_1,...,q_N),~ p=(p_1,...,p_N)$ is described by the Liouville
equation for the function $\rho=\rho (q,p,t)$
\begin{equation}\label{Lio5}
\frac{\partial \rho}{\partial t}=\sum_{i}( \frac{\partial
V(q)}{\partial q_i}\frac{\partial \rho}{\partial
p_i}-\frac{p_i}{m_i}\frac{\partial \rho}{\partial q_i})\,.
\end{equation}
Here summation goes on $ i = 1 ,..., N \, . $ The characteristics
equations  for (\ref {Lio5}) are  Hamilton's equations
\begin{equation}\label{char3}
\dot{q_i}=\frac{\partial H}{\partial p_i},\,\,\dot{
p_i}=-\frac{\partial H}{\partial q_i}\,,
\end{equation}
where the Hamiltonian is
\begin{equation}\label{ham3}
H=\sum_{i}\frac{p_i^2}{2m_i}+V(q)\,.
\end{equation}
Emphasize here again that the Hamilton equations (\ref{char3}) in
the current functional approach to the mechanics do not describe
directly the motion of particles, and they are only the
characteristic equations for the Liouville equation (\ref {Lio5})
which has a physical meaning. The Liouville equation (\ref {Lio5})
can be written as
\begin{equation}\label{Lio6}
\frac{\partial \rho}{\partial t}=\{H,\rho\}\,,
\end{equation}
where the Poisson bracket
\begin{equation}\label{poib}
\{H,\rho\}=\sum_{i}( \frac{\partial H}{\partial q_i}\frac{\partial
\rho}{\partial p_i}-\frac{\partial H}{\partial p_i}\frac{\partial
\rho}{\partial q_i})\,.
\end{equation}

\section{Corrections to Newton's Equations}
In section 3, it was noted that for the free particle in the
functional approach to classical mechanics the averages coordinates
and momenta satisfy the Newton equations. However, when there is a
nonlinear  interaction, then in functional approach corrections  to
the Newton's equations appear.

Consider the motion of a particle along the line in the functional
mechanics. Average value $ \overline {f} $ of the function on the
phase space $ f = f (q, p) $ at time $ t $ is given by the integral
 (\ref{int1})
\begin{equation}\label{defcor}
\overline{f}(t)=<f(t)>=\int f(q,p)\rho (q,p,t)dqdp\,.
\end{equation}
Here  $\rho (q,p,t)$ has the form (\ref{Lio4})
\begin{equation}\label{corr}
\rho (q,p,t)=\rho_0(\varphi_{-t}(q,p))\,.
\end{equation}
By making the replacement of variables,  subject to the invariance
of the Liouville measure, we get
\begin{equation}\label{correcti}
<f(t)>=\int f(q,p)\rho (q,p,t)dqdp=\int f(\varphi_{t}(q,p))\rho_0
(q,p)dqdp\,.
\end{equation}

Let us take
\begin{equation}\label{correc}
\rho_0(q,p)=\delta_{\epsilon}(q-q_0)\delta_{\epsilon}(p-p_0)\,,
\end{equation}
where
\begin{equation}\label{corr1}
\delta_{\epsilon}(q)=\frac{1}{\sqrt{\pi}\epsilon}e^{-q^2/\epsilon^2}\,,
\end{equation}
 $q\in \mathbb{R} , \, \epsilon >0$.

 Let us show that in the limit $\epsilon \to 0$ we obtain
 the Newton (Hamilton) equations:
\begin{equation}\label{correc10} \lim_{\epsilon\to 0}<f(t)>
=f(\varphi_t(q_0,p_0))\,.
\end{equation}

{\bf Proposition 1}. {\it Let the function $f(q,p)$ in the
expression (\ref{defcor})  be continous and integrable, and $\rho_0$
has the form (\ref{correc}). Then}
\begin{equation}\label{correc1}
\lim_{\epsilon\to 0}\int f(q,p)\rho
(q,p,t)dqdp=f(\varphi_t(q_0,p_0))\,.
\end{equation}
{\bf Proof}.
 Functions $\delta_{\epsilon}(q)$ form a
  $\delta$-sequence in $D^{'}(\mathbb{R})$ \cite{Vla}. Hence we obtain
\begin{equation}\label{correc2}
\lim_{\epsilon\to 0}\int f((q,p))\rho (q,p,t)dqdp=\lim_{\epsilon\to
0}\int
f(\varphi_{t}(q,p))\delta_{\epsilon}(q-q_0)\delta_{\epsilon}(p-p_0)=
f(\varphi_t(q_0,p_0))\,,
\end{equation}
that was required to prove.

Now calculate the corrections to the solution of the equation of
Newton. In functional mechanics consider the equation, see
(\ref{Lio})\,,
\begin{equation}\label{Lio2}
\frac{\partial \rho}{\partial t}=-p\frac{\partial \rho}{\partial
q}+\lambda q^2 \frac{\partial \rho}{\partial p}\,.
\end{equation}
Here  $\lambda$ is a small parameter and we set the mass $m=1$. The
characteristic equations have the form of the following Hamilton
(Newton) equations:
\begin{equation}\label{corr3}
\dot{p}(t)+\lambda q (t)^2=0\,,\,\, \dot{q}(t)=p(t)\,.
\end{equation}
Solution of these equations with the initial data
$
q(0)=q,\,\,\dot{q}(0)=p
$
 for small $t$ has the form
\begin{equation}\label{corr5}
(q(t), p(t))=\varphi_t
(q,p)=(q+pt-\frac{\lambda}{2}q^2t^2+...,\,p-\lambda q^2 t+...)
\end{equation}
Use the asymptotic expansion $\delta_{\epsilon}(q)$ in
$D^{'}(\mathbb{R})$ for $\epsilon \to 0$, compare \cite{ALV, PV}:
\begin{equation}\label{corr2}
\delta_{\epsilon}(q)=\delta
(q)+\frac{\epsilon^2}{4}\delta^{''}(q)+...\,,
\end{equation}
then for  $\epsilon \to 0$ we obtain corrections to the Newton
dynamics:
\begin{equation}\label{corr6}
<q(t)>=\int(q+pt-\frac{\lambda}{2}q^2t^2+...)[\delta
(q-q_0)+\frac{\epsilon^2}{4}\delta^{''}(q-q_0)+...]
\end{equation}
$$
\cdot [\delta
(p-p_0)+\frac{\epsilon^2}{4}\delta^{''}(p-p_0)+...]dqdp=q_0+p_0
t-\frac{\lambda}{2}q_0^2t^2 -\frac{\lambda}{4}\epsilon^2t^2\,.
$$
Denoting the Newton solution
$$
q_{\rm Newton}(t)=q_0+p_0 t-\frac{\lambda}{2}q_0^2t^2\,,
$$
we obtain for small $\epsilon, t$ and $\lambda$:
\begin{equation}\label{corr6sm}<q(t)>=
q_{\rm Newton}(t)-\frac{\lambda}{4}\epsilon^2t^2\,.
\end{equation}

Here $-\frac{\lambda}{4}\epsilon^2t^2$  is the correction to the
Newton solution  received within the functional approach to
classical mechanics with the initial Gaussian distribution function.
If we choose a different initial distribution we get correction of
another form.

We have proved

{\bf Proposition 2}.  {\it In the functional approach to mechanics
the first correction at $\epsilon$ to the Newton dynamics for small
$ t$ and $\lambda$ for equation (\ref{corr3}) has the form
(\ref{corr6sm})}.

Note that in the functional approach to mechanics instead of the
usual Newton equation
\begin{equation}\label{New1}
m\frac{d^2}{dt^2}q(t)=F(q)\,,
\end{equation}
where $F(q)$ is a force, we obtain
\begin{equation}\label{Newfunc}
m\frac{d^2}{dt^2}<q(t)>=<F(q)(t)> \,.
\end{equation}
The task of calculating the corrections at $ \epsilon $ for Newton's
equation for mean values is similar to the problem of calculating
semiclassical corrections in quantum mechanics.

\section{Dynamics of a Particle in a Box}

Dynamics of collisionless continuous medium in a box with reflecting
walls is considered by Poincare and in \cite{ Koz1}. This studied
asymptotics of solutions of Liouville equation. In functional
approach to mechanics, we interpret the solution of the Liouville
equation as described the dynamics of a single particle. Here we
consider this model in the classical and  also in the quantum
version for the special case of Gaussian initial data. In particular
we obtain that a single free particle in the box behaves  like a gas
with the Maxwell type distribution.

\subsection {Dynamics of a classical particle in a box}
Consider the motion of a free particle on the interval with the
reflective ends. Using the method of reflections \cite{Vla}, the
solution of the Liouville equation (\ref{free})
$$
 \frac{\partial \rho}{\partial
t}=-\frac{p}{m}\frac{\partial \rho}{\partial q}
$$
on the interval $0\leq q\leq 1$ with the reflective ends we write as
\begin{equation}\label{refl}
\rho (q,p,t)=\sum_{n=-\infty}^{\infty} [\rho_0
(q-\frac{p}{m}t+2n,p)+\rho_0 (-q+\frac{p}{m}t+2n,-p)]\,,
\end{equation}
where it is assumed that the function $\rho_0$ has the Gaussian form
(\ref{gauss}).

One can show that for the distribution for coordinates $ \rho_c
(q,t)=\int \rho (q,p,t)dp $ one gets the uniform limiting
distribution (pointwise limit): $ \lim_{t\to\infty}\rho_c (q,t)=1\,.
$ For the distribution of the absolute values of momenta ($p>0$) $
\rho_a (p,t)=\rho_m (p,t)+\rho_m (-p,t)\,, $ where
$$
\rho_m (p,t)=\int_{0}^{1} \rho (q,p,t)dq\,,
$$
as $t\to\infty$ we get the distribution of the Maxwell type (but not
the Maxwell distribution):
$$
\lim_{t\to\infty}\rho_a
(p,t)=\frac{1}{\sqrt{\pi}b}[e^{-\frac{(p-p_0)^2}{b^2}}
+e^{-\frac{(p+p_0)^2}{b^2}}]\,.
$$
\subsection{Dynamics of a quantum particle in a box}

 The Schrodinger equation for free quantum particle on the interval
  $0\leq x\leq
1$ with reflecting ends has the form
\begin{equation}\label{quansegm}
i\hbar\frac{\partial \phi}{\partial
t}=-\frac{\hbar^2}{2m}\frac{\partial^2\phi}{\partial x^2}
\end{equation}
with the boundary conditions $ \phi (0,t)=0,~~\phi
(1,t)=0,~~t\in\mathbb{R}\,. $ Solution of this boundary problem can
be written as follows:
$$
\phi (x,t)=\sum_{n=-\infty}^{\infty}[\psi(x+2n,t)-\psi(-x+2n,t)]\,,
$$
where $\psi (x,t)$ is some solution of the Schrodinger equation. To
get an observable quantity we have to compute the average value with
the distribution $\sigma (\xi)$ (see Sect.2). Such average values
will demonstrate an irreversible behaviour for large time. If we
choose the function $\psi (x,t)$ in the form, corresponding to the
distribution (\ref{quan2}), then one can show that in the
semiclassical limit for the probability density $|\phi(x,t)|^2$ the
leading term is the classical distribution $\rho_c(x,t)$.

\section{Conclusions}
In this paper the functional formulation of classical mechanics is
considered which is based not on the notion of an individual
trajectory of the particle but on the distribution function on the
phase space.

The fundamental equation of the microscopic dynamics in the
functional approach is not the Newton equation but the Liouville
equation for the distribution function of a single particle.
Solutions of the Liouville equation have the property of
delocalization which accounts for irreversibility. It is shown that
the Newton equation in this approach appears as an approximate
equation describing the dynamics of the average values of the
positions and momenta for not too long time intervals. Corrections
to the Newton equation are computed.

If we accept the functional approach to classical mechanics then
both classical and quantum mechanics contain fundamental randomness.
It requires a reconsideration of the usual interpretation(s) of
quantum mechanics. Some remarks on these questions we made in
Sect.2. Interesting problems related with applications of the
functional mechanics to statistical mechanics, field theory,
singularities in cosmology and black holes we hope to consider in
further works.

\section{Acknowledgements}
I am grateful to many colleagues for valuable discussions. Some
results of this paper were presented  at the QTRF5 (Vaxjo, Sweden)
and QBIC2009 (Tokyo) conferences. The work is partially supported by
grants NS-3224.2008.1, RFBR 08-01-00727-a, 09-01-12161-ofi-m, AVTSP
3341, DFG Project 436 RUS 113/951.

 \end{document}